\documentclass[aps, prb, reprint, twocolumn, superscriptaddress]{revtex4-1}

\usepackage{lipsum}
\usepackage{multirow}
\usepackage{graphicx}
\usepackage{longtable}
\usepackage[utf8]{inputenc}
\usepackage[T1]{fontenc}
\usepackage{epstopdf}
\usepackage{textcomp}
\usepackage{color}
\usepackage[cmyk,dvipsnames]{xcolor} % For obvious reasons
\usepackage{amsbsy}
\usepackage{amsmath}
\usepackage{amssymb}
\usepackage{amsfonts}
\usepackage{dsfont}
\usepackage{mathtools}
\usepackage{braket}
\usepackage{array}
\usepackage{gensymb}
\usepackage{placeins}
\usepackage{dashrule}
\usepackage{xcolor}
\usepackage{booktabs}
\usepackage{hyperref}
\usepackage{float}
\usepackage{braket}
\usepackage{ulem}

\definecolor{colorA}{rgb}{0, 0, 1}
\definecolor{colorB}{rgb}{0.5, 0, 0.9}

\makeatletter
  \def\my@tag@font{\normalsize}
  \def\maketag@@@#1{\hbox{\m@th\normalfont\my@tag@font#1}}
  \let\amsmath@eqref\eqref
  \renewcommand\eqref[1]{{\let\my@tag@font\relax\amsmath@eqref{#1}}}
\makeatother

\allowdisplaybreaks

\begin{document}

\title{Geometry and Symmetry in Skyrmion Dynamics}

\author{Vladyslav~M.~Kuchkin}
\email{v.kuchkin@fz-juelich.de}
\affiliation{Peter Gr\"unberg Institute and Institute for Advanced Simulation, Forschungszentrum J\"ulich and JARA, 52425 J\"ulich, Germany}
\affiliation{Department of Physics, RWTH Aachen University, 52056 Aachen, Germany}

\author{Ksenia~Chichay}
%\email{ks.chichay@gmail.com}
\affiliation{Immanuel Kant Baltic Federal University, 236041 Kaliningrad, Russia}

\author{Bruno~Barton-Singer}
\affiliation{Maxwell Institute for Mathematical Sciences and Department of Mathematics, Heriot-Watt University, Edinburgh, EH14 4AS, UK}

\author{Filipp~N.~Rybakov}
% \email{f.n.rybakov@gmail.com}
\affiliation{Department of Physics, KTH Royal Institute of Technology, Stockholm, SE-10691 Sweden}

\author{Stefan Bl\"ugel}
 \affiliation{Peter Gr\"unberg Institute and Institute for Advanced Simulation, Forschungszentrum J\"ulich and JARA, 52425 J\"ulich, Germany}

\author{Bernd J. Schroers}
\affiliation{Maxwell Institute for Mathematical Sciences and Department of Mathematics, Heriot-Watt University, Edinburgh, EH14 4AS, UK}

\author{Nikolai~S.~Kiselev}
 %\email{n.kiselev@fz-juelich.de}
 \affiliation{Peter Gr\"unberg Institute and Institute for Advanced Simulation, Forschungszentrum J\"ulich and JARA, 52425 J\"ulich, Germany}

\date{\today}

\begin{abstract}
%127 words 
The uniform  motion of chiral magnetic skyrmions induced by a  spin-transfer torque  displays an  intricate dependence on the skyrmions' topological charge and  shape.
We  reveal surprising   patterns in this dependence through simulations of the Landau-Lifshitz-Gilbert equation with Zhang-Li torque and explain them  through a geometric analysis of Thiele's equation.
In particular, we show that the velocity distribution of  topologically non-trivial skyrmions depends on their symmetry:  it is a single circle for skyrmions of high symmetry and a family of circles for low-symmetry configurations.
We also show that the velocity of the topologically trivial skyrmions, previously believed to be the fastest objects, can be surpassed, for instance, by antiskyrmions. The generality of our approach suggests the validity of our results for exchange frustrated magnets, bubble materials, and others. 
\end{abstract}

\maketitle

\begin{figure}[t]
\centering
\includegraphics[width=7.5cm]{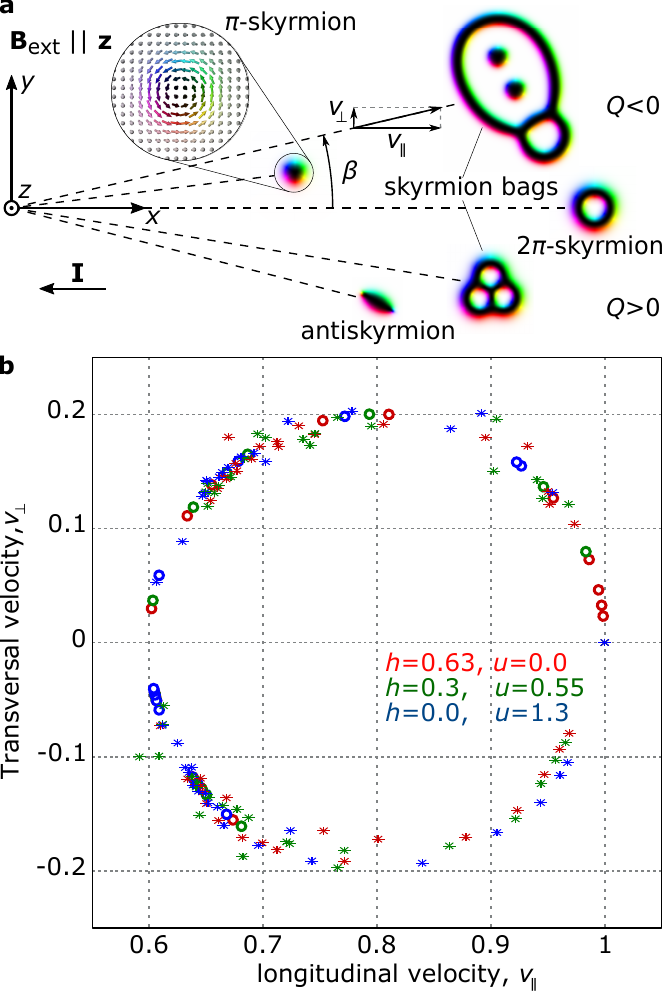}
\caption{\small {\bf Representative example for skyrmions velocity distribution.} 
\textbf{a} Schematic representation for  the trajectories of chiral skyrmions moving under electric current $\mathbf{I}$. 
$\beta$ is the skyrmion deflection angle.
The inset showing the spin texture for $\pi$-skyrmion explains the color code.
The antiskyrmion is provided as an example of a skyrmion with chiral kinks~\cite{Kuchkin_20ii}.
\textbf{b} The velocity distribution for a wide diversity of skyrmions obtained in micromagnetic simulations at different values of the external magnetic field $h$ and anisotropy $u$. 
The velocities represented by the circle symbols lie on a perfect circle with the centre at $v_\parallel=0.8$, $v_\perp=0$ irrespective of $h$ and $u$.
The velocities marked by the stretched star symbols lie in the close vicinity to that circle, for details see main text.
The velocities are given in  reduced units with respect to the velocity of $2\pi$-skyrmion. The simulations are performed at realistic values of the Gilbert damping $\alpha=0.06$ and the degree of non-adiabaticity, $\xi=0.1$.  
\label{Fig1}
}
\end{figure}

% \section*{Introduction}
%
Many models of two-dimensional (2D) chiral magnets allow the existence of statically stable topological solitons -- localized magnetic textures possessing particle-like properties~\cite{Bogdanov_89}.
Nowadays, it is common to refer to these objects as chiral magnetic skyrmions. For convenience, we consider skyrmions to include localised configurations with zero topological charge.
The axisymmetric solutions representing vortex-like spin textures as $\pi$-skyrmion, or more generally $k\pi$-skyrmions  (Fig.~\ref{Fig1}\textbf{a}),  have been  much studied since the  pioneering works by Bogdanov, Yablonskii, and Hubert \cite{Bogdanov_89, Bogdanov_1994,Bogdanov_1994JMMM, Bogdanov_99}. 
Most of the previously published papers discussing different properties of chiral magnetic skyrmions are devoted to such $k\pi$-skyrmions. 
Only recently, new classes of skyrmion solutions with diverse morphology and arbitrary topological charge have been reported.~\cite{Rybakov_19, Foster_19, Kuchkin_20i, Kuchkin_20ii,
Barton-Singer_20}
The static properties of so-called skyrmion bags\cite{Rybakov_19, Foster_19} and skyrmions with chiral kinks\cite{Kuchkin_20i, Kuchkin_20ii} belonging to these newly discovered classes of skyrmions are now quite  well understood.

However, there are only a few  papers  considering the dynamics of such skyrmions \cite{Kind_21, Zeng_20}, and a systematic study of their dynamical properties is missing. 
Here, we study the uniform motion of skyrmions induced by the  Zhang-Li spin-transfer torque \cite{ZhangLi}. We use numerical micromagnetic simulations based on the Landau-Lifshitz-Gilbert equation and a semi-analytic method based on the Thiele approach. 
In the presence of an electric current, skyrmions generally move in the direction opposite to the current vector $\mathbf{I}$, but the longitudinal  and transverse velocity components $v_{\parallel}$  and  $v_{\perp}$  depend on the  skyrmion (Fig.\ref{Fig1}\textbf{a}).
Here we find, using micromagnetic simulations,  that the velocity distribution has remarkable geometrical features and is concentrated in a ring-like  region in the $(v_{\parallel},v_{\perp})$ plane,  shown in (Fig.\ref{Fig1}\textbf{b}).
The investigation of this intriguing phenomenon is the main subject of the present paper. 

We show that the skyrmion velocity distribution can be understood by splitting all skyrmions into high-symmetry, low-symmetry and topologically trivial ones.
Irrespective of the magnetic field and anisotropy, the velocities of high-symmetry skyrmions, for instance axially symmetric $k\pi$-skyrmions, always lie on a circle. 
The radius of that circle depends exclusively on the current density and the internal parameters of the system, such as Gilbert damping and the coefficient of nonadiabaticity of the electric current.
The low-symmetry skyrmions exhibit an even more intriguing behavior -- their velocities depend on the skyrmion orientation with respect to the current direction. 
Interestingly, the velocity distribution for each low-symmetry skyrmion also represents a circle of, generally,  smaller radius. 
Moreover, it is shown that the variation of the internal parameters of the system can lead to the degeneration of all  those circles into one point when all skyrmions move along one trajectory with the same velocity.

% \section*{Results}

\vspace{0.25cm}
\noindent
\textbf{Results}

\noindent
\textbf{Micromagnetic simulations}.
We consider the 2D micromagnetic model for a chiral magnet containing three main energy terms:
\begin{align}
\mathcal{E}=\int \{w_\mathrm{ex}(\textbf{n}) + w_\mathrm{D}(\textbf{n})+w_\mathrm{U}(\textbf{n})\}\,  l\mathrm{d}x\mathrm{d}y,\label{Etot}
\end{align}
where $\textbf{n}=\textbf{M}/M_\mathrm{s}$ is the magnetization unit vector field uniform across the film thickness, $l$,  $M_\mathrm{s}$ is the saturation magnetization, $w_\mathrm{ex}=\mathcal{A}\left|\nabla\mathbf{n}\right|^{2}$ is the Heisenberg exchange interaction
and $w_\mathrm{U}=-M_\mathrm{s}\mathbf{B}_\mathrm{ext} \cdot\mathbf{n}-\mathcal{K} n_\mathrm{z}^{2}$ 
is the potential term containing the Zeeman interaction and the easy-axis/easy-plane anisotropy.

The Dzyaloshinskii-Moriya interaction~\cite{Dzyaloshinskii,Moriya} (DMI) term $w_\mathrm{D}(\mathbf{n}) = \mathcal{D} w(\mathbf{n})$ 
is defined by combinations of Lifshitz invariants, $\Lambda_{ij}^{(k)}\!=\! n_i\partial_k n_j\!-\!n_j\partial_k n_i$.
The results presented below are valid for classes of chiral magnets of different crystal symmetries with: N\'{e}el-type modulations~\cite{Romming_13,Kez_15,Romming_15} where 
$w(\mathbf{n})\!=\!\Lambda_{xz}^{(x)}\!+\! \Lambda_{yz}^{(y)}$,
D$_{2\mathrm{d}}$ symmetry~\cite{Nayak_17} where 
$w(\mathbf{n}) \!= \!\Lambda_{zy}^{(x)}\!+\! \Lambda_{zx}^{(y)}$,  and 
Bloch-type modulations~\cite{ElisaGiovanni}
where  $w(\mathbf{n})\! =  \!\Lambda_{zy}^{(x)}\!+\! \Lambda_{xz}^{(y)}\!$. Without loss of  generality, the latter is used by default in our calculations.
Here we assume that the external magnetic field is uniform and always perpendicular to the plane of the film, $\textbf{B}_\mathrm{ext}\parallel\textbf{e}_\mathrm{z}$.
Introducing the characteristic size of chiral modulations $L_\mathrm{D}=4\pi\mathcal{A}/\mathcal{D}$ and the critical magnetic field $B_\mathrm{D}=\mathcal{D}^2/(2 M_\mathrm{s}\mathcal{A})$ one can  
reduce the number of independent parameters to two, $h=B_\mathrm{ext}/B_\mathrm{D}$ and $u = \mathcal{K}/(M_\mathrm{s} B_\mathrm{D})$ corresponding to dimensionless external
magnetic field and anisotropy, respectively.

Skyrmion motion can be caused by different stimuli, e.g.\ the gradient of the internal or external parameters and spin-orbit or spin-transfer torques. Here we consider the particular case of the Zhang-Li\cite{ZhangLi} spin-transfer torque. 
The Landau-Lifshitz-Gilbert (LLG) equation~\cite{Landau_Lifshitz} in this case has the following form
\begin{align}
 & \dfrac{\partial\mathbf{n}}{\partial t}=-\gamma\mathbf{n}\times\mathbf{H}_\mathrm{eff}+\alpha\mathbf{n}\times\dfrac{\partial\mathbf{n}}{\partial t} -\textbf{T}_\mathrm{ZL},\label{LLG}
\end{align}
where $\gamma$ is the gyromagnetic ratio, $\alpha$ is the Gilbert damping, and the effective field   $\mathbf{H}_\mathrm{eff}=-\dfrac{1}{M_\mathrm{s}}\dfrac{\delta\mathcal{E}}{\delta\mathbf{n}}$ is defined by variation of the total energy $\mathcal{E}$. 
The last term in \eqref{LLG} is the Zhang-Li torque:
\begin{align}
\textbf{T}_\mathrm{ZL}= \mathbf{n}\!\times\!\left[\mathbf{n}\times\left(\mathbf{I}\cdot\nabla\right)\mathbf{n}\right]+\xi\,\mathbf{n}\times\left(\mathbf{I}\cdot\nabla\right)\mathbf{n},
\label{T_ZL}
\end{align} 
where the vector $\mathbf{I}=\mathbf{j}{\mu_\mathrm{B}p}(1\!+\!\xi^{2})^{-1}(e M_\mathrm{s})^{-1}$ is proportional to the current density $\mathbf{j}$,
$\xi$ is the degree of non-adiabaticity\cite{Malinowski},  $p$ is the polarization of the spin current, $\mu_\mathrm{B}$ is the Bohr magneton and  $e$ is the electron charge. 
Note that $I=I(\xi)$, thereby, assuming that $I=\mathrm{const}$ for varying $\xi$  requires  the current density $\mathbf{j}$  to change.

We study the solutions of equation~(\ref{LLG}) corresponding to uniform motion of magnetic skyrmions.
Using different values of the external field $h$ and anisotropy $u$  in  micromagnetic simulations for a large diversity of skyrmions (see Methods section), we obtain the striking velocity distribution shown in  Figure~\ref{Fig1}\textbf{b}. 
To explain the striking  circular shape of this distribution we employ  analytical methods described below.

\vspace{0.25cm}
\noindent
\textbf{Equation of motion}.
The uniform motion of magnetic textures is well described by Thiele's equation \cite{Thiele_73} which can be derived from the  LLG equation \eqref{LLG} and,  for the Zhang-Li spin-transfer torque~\eqref{T_ZL},  has the following form~\cite{Komineas}:
\begin{align}
	-Q\,\mathbf{e}_\mathrm{z}\!\times\!\left(\mathbf{V}+\mathbf{I}\right)-\Gamma \left(\alpha\mathbf{V}+\xi\mathbf{I}\right) =0,
	\label{Thiele}
\end{align}
where $\mathbf{V}=(V_{\mathrm{x}}, V_{\mathrm{y}})^\mathrm{T}$ is the velocity vector of the skyrmion moving as a rigid object, i.e.\ $\textbf{n}(\mathbf{r},t)=\textbf{n}_0(\mathbf{r}-\mathbf{V}t)$.
The two essential parameters in \eqref{Thiele} are the topological charge~\cite{Q_note}
\begin{equation}
Q = \dfrac{1}{4\pi}\int\!  \mathbf{n}\cdot \left(\partial_\mathrm{x}\mathbf{n}\times\partial_\mathrm{y}\mathbf{n}\right) \,\mathrm{d}x\mathrm{d}y,
\label{Qint}
\end{equation}
and the dissipation tensor $\Gamma$, whose components  are given by the following integrals~\cite{Malozemoff_79, Malozemoff_note}   
 \begin{equation}
\Gamma_{ij}
=\dfrac{1}{4\pi}\ \int \left(\partial_i\mathbf{n} \cdot \partial_j\mathbf{n} \right) \mathrm{d}x\mathrm{d}y, \quad i,j=x,y.
\label{dissipation}
\end{equation}

Thiele's equation  \eqref{Thiele} has  a simple algebraic form, but to solve it with respect to $\mathbf{V}$ one has to know the skyrmion magnetization profile $\mathbf{n}_0$ and then calculate the integrals in \eqref{Qint} and \eqref{dissipation}. 
In general, $\mathbf{n}_0$ is any configuration consistent with the derivation of Thiele's equation

and LLG equation \eqref{LLG}. 
However, previous numerical studies~\cite{Komineas} have shown that uniform motion has only a secondary effect on the skyrmion profile.
Thus, it is natural to expect that the tensor $\Gamma$ can be calculated with good accuracy if a static equilibrium  configuration is chosen as the skyrmion magnetization profile in~(\ref{dissipation}).
Accordingly, to calculate the tensor $\Gamma$ we use solutions 
found by numerical minimization of \eqref{Etot} by means of conjugate gradient method and fourth-order finite-difference scheme implemented in  the Excalibur code~\cite{Excalibur}.
This semi-analytical approach based on solutions of Thiele's equation~\eqref{Thiele} with static solutions for $\mathbf{n}$ shows very good agreement with the results of direct micromagnetic simulations based on LLG equation~\eqref{LLG}. 
That means that to understand the physical nature of the circular shape of the velocity distribution observed in a numerical experiment (Fig.~\ref{Fig1}\textbf{b}) one can rely on the analysis of Thiele's equation.

\begin{figure}[t]
\centering
\includegraphics[width=7.5cm]{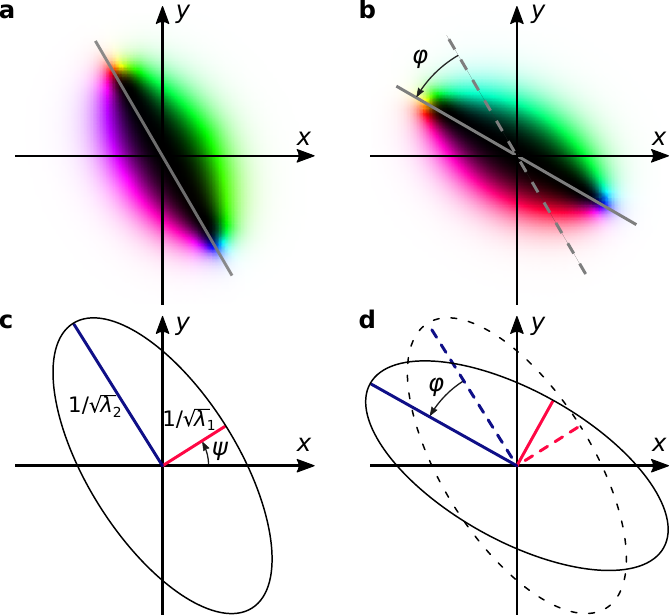}
\caption{\small {\bf Illustration of the rotational  zero energy mode}. The spin texture of an antiskyrmion before (\textbf{a}) and after (\textbf{b}) the rotation by an angle $\varphi$ according to \eqref{Rn}. The spin textures in \textbf{a} and \textbf{b} are represented by  the  standard color code explained  in Fig.~\ref{Fig1}\textbf{a}. The  ellipses in \textbf{c} and \textbf{d} are the geometrical representations of $2\times 2$ matrix of  $\Gamma$ computed for the spin textures depicted in \textbf{a} and \textbf{b}, respectively. 
\label{FigRot}
}
\end{figure}

\begin{figure*}[ht]
\centering
\includegraphics[width=16.5cm]{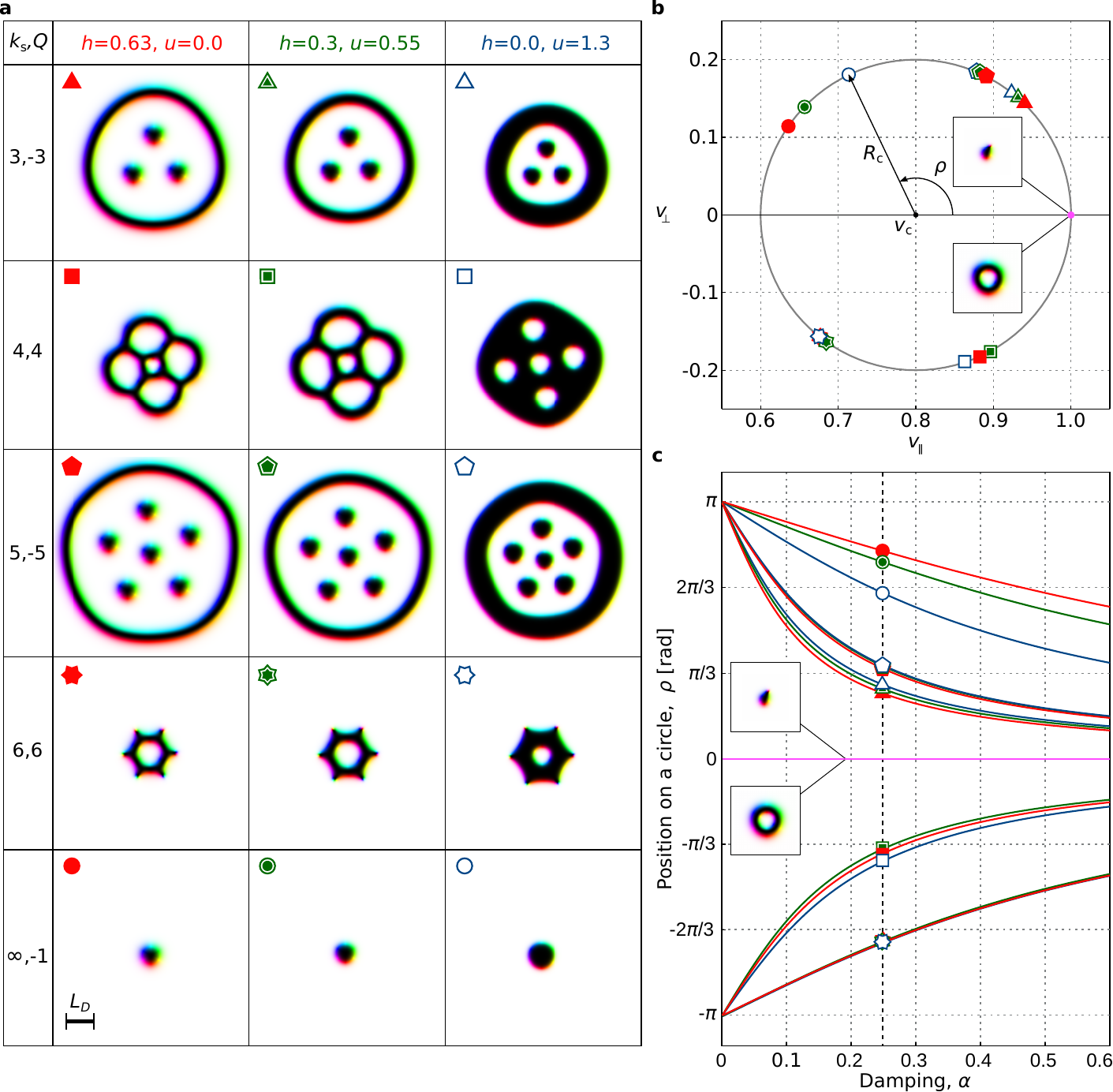}
\caption{\small \textbf{The velocity circle for high-symmetry skyrmions}. 
(\textbf{a}) The set of high-symmetry skyrmions at different external fields, $h$ and anisotropies, $u$. The parameter $k_\mathrm{s}$ in the first column stands for the order of the rotational symmetry of the skyrmion and $Q$ is the topological charge \eqref{Qint}.
The $\pi$-skyrmion in the last row has axial symmetry, $k_\mathrm{s} = \infty$. 
(\textbf{b}) The velocities of skyrmions for particular symmetry and particular values of $h$ and $u$ are depicted with the corresponding symbols.
The velocity of topologically trivial solutions, e.g.\ $2\pi$-skyrmion and chiral droplet (see insets), is marked with a magenta circle, $v_{\parallel}=1$, $v_{\perp}=0$.
The calculations based on semi-analytical approach are performed for $\alpha=1/4$, $\xi/\alpha = 5/3$.
The grey circle corresponds to~\eqref{big_circle}.
(\textbf{c})~The change of the distribution of velocities on the circle in terms of angle $\rho$ for the solitons depicted in \textbf{a}  as a function of  $\alpha\in\left(0,0.6\right]$, for fixed $\xi/\alpha = 5/3$. The dashed line corresponds to damping parameter $\alpha=1/4$ as in \textbf{b}. 
}
\label{big_circle}
\end{figure*}

\vspace{0.25cm}
\noindent
\textbf{Rotational symmetry of skyrmions}.
Although not obvious at first, 
the key to understanding skyrmion dynamics is  the relationship between the symmetry of skyrmions and the  parameters in Thiele's equation. 
Let us consider the transformation representing the rotation of the whole spin texture about the axis normal to the plane:
\begin{align}
&\mathbf{n}^\prime (\mathbf{r}) = \mathcal{R}(\varphi)\,\mathbf{n} \!\left(\mathcal{R}(-\varphi)\mathbf{r}\right), \label{Rn}
\end{align}
where $  \mathcal{R}(\varphi)$ is the   $3\times 3$  matrix for a rotation by $\varphi$  about the $z$-axis. For the  Hamiltonian~\eqref{Etot} with Bloch or Neel DMI,  the rotation~(\ref{Rn}) represents a zero-energy mode, see Fig.~\ref{FigRot}\textbf{a}, \textbf{b}.
When the transformation \eqref{Rn} with $\varphi=2\pi/k$ is trivial for some positive integers $k$, so that ${\mathbf{n}^\prime = \mathbf{n}}$ (possibly up to translation), 
we say that the spin texture has a rotational symmetry of order $k_\mathrm{s}=\mathrm{max}(k)$.
For axially symmetric skyrmions, e.g.\ $\pi$-skyrmion, the invariance holds for any $k$ and we write  $k_\mathrm{s}=\infty$, see e.g. Fig.~\ref{big_circle}\textbf{a}.

The skyrmions possessing rotational symmetry of order  $k_\mathrm{s}>2$  have the property that  the  response velocity determined by  Thiele's equation \eqref{Thiele} is invariant under the rotation \eqref{Rn}  of a skyrmion  by an arbitrary angle $\varphi$, while for skyrmions with $k_\mathrm{s}=1$ or 2, 
the response velocity in general depends on the  rotation angle.
This statement can be proven as follows.
Inserting \eqref{Rn} into \eqref{Qint} and \eqref{dissipation} one can show that for any configuration localised in space, the topological charge $Q$ is invariant under such rotations.
On other hand, the dissipation tensor transforms according to $ \Gamma^\prime = R(\varphi)\, \Gamma \, R(-\varphi)$,
where $R(\varphi)$ is the $2\times 2$ matrix for a (mathematically positive) rotation by $\varphi$ in the plane. This transformation law  has the  more convenient representation
\begin{equation}
\mathbf{S}^\prime= R(2\varphi) \mathbf{S},
\label{trS}
\end{equation}
where $\mathbf{S}$ is a 2D vector, 
$
\mathbf{S}=\left( \Gamma_\mathrm{xx}-\Gamma_\mathrm{yy},\, 2\Gamma_\mathrm{xy} \right)^\mathrm{T}.
$
If the spin texture is invariant under rotations by angles $\varphi=2\pi/k_\mathrm{s}$ ($\mathbf{n}^\prime=\mathbf{n}$) it follows that $\Gamma^\prime=\Gamma$ and $\mathbf{S}^\prime=\mathbf{S}$ for those angles. 
For skyrmions with $k_\mathrm{s}=1$ or 2, this condition is satisfied automatically, since   $R\left(4\pi/k_\mathrm{s}\right)$
is the identity matrix, $R= \mathrm{id}$.
For such spin configurations, the components of the vector $\mathbf{S}$, and $\Gamma$ tensor may, strictly speaking, take any value. 
For skyrmions with rotational symmetry $k_\mathrm{s}>2$, on the other hand,  it follows that $\mathbf{S}$ must be a zero vector,  and  thus $\Gamma_\mathrm{xx}=\Gamma_\mathrm{yy}$, $\Gamma_\mathrm{xy}=0$,  meaning that $\Gamma$ is proportional to an identity matrix,  $\Gamma=\frac{1}{2}(\Gamma_\mathrm{xx}+\Gamma_\mathrm{yy})\,\mathrm{id}=\frac{1}{2}\mathrm{Tr}(\Gamma)\,\mathrm{id}$.
It follows from \eqref{trS} that for such skyrmions the dissipation tensor and, as a result, the velocities determined by \eqref{Thiele} are indeed invariant under rotations \eqref{Rn} by an arbitrary angle $\varphi$.
Motivated by this proof,  we distinguish  topologically non-trivial skyrmions by their symmetry. 
We refer to  skyrmions with $k_\mathrm{s}=1$ or 2 as \textit{low-symmetry skyrmions}
and to skyrmions with $k_\mathrm{s}>2$ as  \textit{high-symmetry skyrmions}.
The dynamical properties of skyrmions with $Q=0$ do not depend on the $\Gamma$ tensor at all, and we refer to them   as a third class of  \textit{topologically trivial skyrmions}.

We now provide an  analysis of Thiele's equation which  shows  how the skyrmions' symmetry influences their dynamics. 

\vspace{0.25cm}
\noindent
\textbf{The hidden geometry of  Thiele's equation}.
For a given  current  $\mathbf{I}$, Thiele's equation \eqref{Thiele} determines the dependence of the  velocity   on the topological charge, the  dissipation tensor and the material parameters $\alpha$ and $\xi$. This dependence has a surprisingly rich geometry  which does not appear to have been studied in the literature. 
For our discussion we  rescale the velocity by the speed of skyrmionium~\cite{Komineas} $V_0=\xi|\mathbf{I}|/\alpha$ and define $\mathbf{v} =- \mathbf{V}/V_0$. 
We  also introduce an  oriented orthonormal basis $(\mathbf{e}_\parallel,\mathbf{e}_\perp)$ adapted to the direction of the current by choosing  $\mathbf{e}_\parallel$ to be   anti-parallel to   $\mathbf{I}$.
The  geometrical beauty  of  Thiele's equation becomes evident when the dissipation tensor  is expressed in terms of its real and positive eigenvalues, see  Fig.~\ref{FigRot}. 
As a symmetric  and positive $2\times 2$  matrix,  $\Gamma$ can be brought into diagonal form by conjugation with a rotation matrix. Denoting the  real eigenvalues by  $\lambda_1\geq \lambda_2>0$ and the rotation angle by $\psi$,  we   have the parametrisation of  $\Gamma$ as 
\begin{equation}
\Gamma   = 
 R(\psi)  \begin{pmatrix} \lambda_1  & 0 \\ 0 &\lambda_2\end{pmatrix}
 R(-\psi). 
\end{equation}
The angle $\psi$ is only defined when $\lambda_1>\lambda_2$ and only takes values in $[0,\pi)$. It parametrises the unoriented direction of the eigenvector for $\lambda_1$ relative to the $x$-axis.   Under rotation of a configuration by $\varphi$ according to  \eqref{Rn}, the angle $\psi$ shifts to $\psi +\varphi$, see again Fig.~\ref{FigRot}.

Encoding the geometric mean  and ratio of the eigenvalues of $\Gamma$  into the parameters 
 \begin{equation}
\rho  =2 \tan^{-1}\left(\frac{Q}{\alpha \sqrt{\lambda_1\lambda_2}}\right), \quad  \vartheta =\ln \sqrt{\frac{\lambda_1}{\lambda_2}},
\label{rhovartheta}
\end{equation}
we show in the Methods section  that the general solution of Thiele's equation can usefully  be written as 
 \begin{align}
\mathbf{v}&= \left(v_\mathrm{c}+ R_\mathrm{c} \cos\rho\right)\mathbf{e}_\parallel \nonumber \\
& -R_\mathrm{c}  \sin\rho \left(\cosh \vartheta   - \sinh \vartheta\, R(2\psi) P\right)\mathbf{e}_\perp,
\label{keymaster}
\end{align}
where  $P$ is the matrix 
for the reflection on
the $x$-axis and the parameters
\begin{equation}
 v_\mathrm{c} = \frac{ \xi+\alpha}{2\xi}, \quad  R_\mathrm{c}= \dfrac{ \xi-\alpha}{2\xi},
 \label{circleparameters}
\end{equation} 
are determined by the Gilbert damping and the degree of non-adiabaticity.
 
The  formula \eqref{keymaster}  captures the geometry referred to in our title  and provides the key to understanding the ring-like velocity distribution depicted in Fig.~\ref{Fig1}\textbf{b}. 
When $Q=0$, the velocity takes the  single  value $\mathbf{v}=\mathbf{e}_\parallel$,  regardless of the form of the dispersion tensor, but when $Q\neq 0$ several circular orbits appear in the velocity plane.
The velocities of  skyrmions with  $\vartheta=0$ ($\lambda_1=\lambda_2$) but 
 different values of $\rho$  lie on a circle with radius $R_\mathrm{c}$ and centre $v_\mathrm{c}$.
In terms of the components of $\mathbf{v}$ with respect to the basis $(\mathbf{e}_\parallel,\mathbf{e}_\perp)$, we have 
\begin{equation}
\lambda_1 =\lambda_2 \Rightarrow (v_\parallel -v_\mathrm{c})^2 + v_\perp^2 =R_\mathrm{c}^2.
\label{vcircle}
\end{equation}

For a skyrmion with  $\vartheta >0$ ($\lambda_1>\lambda_2$)
and some fixed value of  $\rho$,
 the velocity  sweeps out a circle as $\psi$ varies, i.e.\ when we physically rotate the skyrmion configuration, as illustrated in  Fig.~\ref{FigRot}. This circle is traversed {\it twice} when we rotate a configuration through $2\pi$; it  has  centre coordinates
 \begin{equation}
     (v^0_\parallel,v^0_\perp)=(v_\mathrm{c}+R_\mathrm{c}\cos\rho,  - R_\mathrm{c}\sin\rho \cosh\vartheta) 
     \label{psicirclecentre}
 \end{equation}
 and   radius 
 \begin{equation}
R_0=R_\mathrm{c}|\sin\rho |\sinh \vartheta= R_c\frac{\alpha| Q| (\lambda_1-\lambda_2)}{\alpha^2\lambda_1\lambda_2 +Q^2}.
\label{psicircleradius}
 \end{equation}
As we shall explain below, these circles tend to have centres close to the circle \eqref{vcircle}, and radii smaller than $R_c$, leading to the ring-like distribution centred on the circle \eqref{vcircle} which we see in Fig.~\ref{Fig1}\textbf{b}.
More generally  varying $\rho$ in \eqref{keymaster} while fixing  $\vartheta>0$ and $\psi$ generates half-ellipses. This can be seen directly from  \eqref{keymaster} whose dependence of $\rho$ is that found in  elliptical Lissajous figures, but is explained in  more detail in  the Methods section.

\vspace{0.25cm} 
\noindent
\textbf{High-symmetry skyrmions}.
Figure~\ref{big_circle} shows representative examples of high-symmetry skyrmions and their corresponding velocity distribution calculated with the semi-analytical approach described above and verified by direct micromagnetic simulation  of  the LLG equation. 
Since high-symmetry skyrmions necessarily have a dissipation tensor with equal eigenvalues, we compare the results of the simulation with the prediction of the general solution \eqref{keymaster} for $\vartheta=0$.
For fixed $\xi/\alpha$ and irrespective of the external magnetic field, $h$, and anisotropy, $u$, which significantly change the shape and size of the skyrmions (Fig.~\ref{big_circle}\textbf{a}), the velocities of all high-symmetry skyrmions are restricted to the circle  \eqref{vcircle}.
The velocities of skyrmions with  $Q<0$ and $Q>0$  occupy half of the circle in the upper or lower half-plane depending on $\textrm{sign}(\xi-\alpha)$.

The position of an individual skyrmion on the circle, parametrised  by the angle $\rho$,  can be linked to experimentally measurable deflection angle\cite{Malozemoff_79}, $\beta=\arctan\left(v_{\perp}/v_{\parallel}\right)$. It follows from the law of sines that 
\begin{equation}
\tan\beta = \dfrac{\sin\rho\sin\beta_\mathrm{max}}{1+\cos\rho\sin\beta_\mathrm{max}},\label{rho}
\end{equation}
where $\beta_\mathrm{max}=\arcsin{\left(R_\mathrm{c}/v_\mathrm{c}\right)}$ is the maximal deflection angle for high-symmetry skyrmions.
Fig.~\ref{big_circle}\textbf{c} shows the variation  of the skyrmions position on the circle
for each of skyrmions depicted in Fig.~\ref{big_circle}\textbf{a}.  
The angle $\rho$ is shown as function of $\alpha$ varying in the range $\left(0,3/5\right]$ for fixed ratio of $\xi/\alpha = 5/3$.

\begin{figure}
\centering
\includegraphics[width=7.5cm]{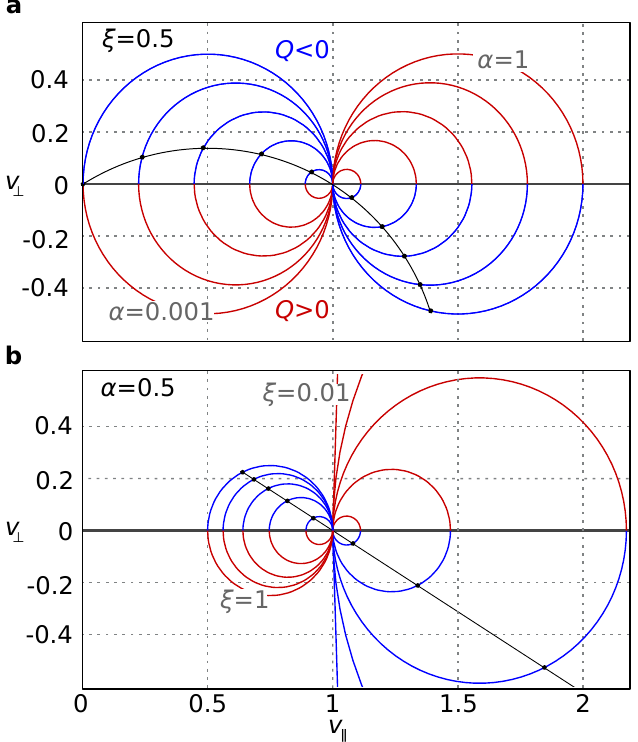}
\caption{\small \textbf{Transformation of the velocity circle for  high-symmetry skyrmions}.
(\textbf{a}) The transformation of the circle \eqref{vcircle} under varying the  Gilbert damping, $\alpha$, at fixed 
degree of non-adiabaticity $\xi=0.5$. The circles correspond to $\alpha$ in the interval $\left[0.001,1\right]$ with equidistant step of 0.111.
(\textbf{b}) The transformation of the circle \eqref{vcircle} under varying degree of non-adiabaticity, $\xi$, at  fixed $\alpha=0.5$. The circles correspond to $\xi$ in the interval $\left[0.01,1\right]$ with equidistant step 0.11.
Black dots in \textbf{a} and \textbf{b} are the velocities of ordinary $\pi$-skyrmion. 
For $\alpha=\xi=0.5$ the velocity circle degenerate into the point, $v_{\parallel}=1,v_{\perp}=0$ which is the common point for all circles.
Red and blue colors of the half-circles correspond to $Q>0$ and $Q<0$ respectively.
Note, the colors are inverted depending on the sign of $1-\alpha/\xi$.
\label{big_circle_transform}
}
\label{AlphaXi}
\end{figure}

Note that the velocity for topologically trivial solitons, e.g.\ $2\pi$-skyrmion (skyrmionium) and the chiral droplet~\cite{Kuchkin_20ii,SisodiaKomineas} (see insets in Fig.~\ref{big_circle}\textbf{b}), are restricted to a single point on the circle, $v_{\parallel}=1, v_{\perp}=0$. These solutions do not belong to the class of high-symmetry skyrmions and are presented only for comparison. 
For the physically realistic case of $\xi>\alpha$,  skyrmions with $Q=0$ have  both a higher speed and a higher velocity component  $v_\parallel$ than any  high-symmetry skyrmion.
With decreasing $\alpha$ the speeds of high-symmetry skyrmions decrease and as $\alpha \rightarrow 0$ we find  $\rho\rightarrow -\textrm{sign}(Q)\pi$, in  accordance with  \eqref{rhovartheta}.
With increasing $\alpha$, the skyrmion  speeds  increase and in the limit $\alpha\gg1$  their velocities approach that of skyrmionium. For the circle parameter $\rho$, this means $ \rho\rightarrow 0$ again in  accordance with \eqref{rhovartheta}. 

\vspace{0.25cm}
\noindent
\textbf{Dependence on $\alpha$ and $\xi$}.
An important aspect of the velocity formula  \eqref{keymaster} is the dependence of $R_\mathrm{c}$ and $v_\mathrm{c}$ in  \eqref{circleparameters} on $\alpha$ and $\xi$.
Figure~\ref{big_circle_transform} illustrates the evolution of the velocity circle for different ratios $\xi/\alpha$.
In contrast to the case of $\xi>\alpha$, for $\xi<\alpha$, $V_0$ becomes the  lower bound of the speed  of  high-symmetry skyrmions. 
In the case of $\xi=\alpha$, the velocity circle degenerate into a single point, meaning that all skyrmions move with the  same velocity $\mathbf{V}_0=V_0\mathbf{e}_\parallel$,  without deflection.
This shows the fundamental and dual physical significance  of the velocity  $\mathbf{V}_0$ as both the velocity of topologically trivial skrymions  for  any value of  the parameters $\alpha$ and $\xi$,  and as the velocity of all skyrmions under the special condition $\alpha=\xi$. 
The sign for the skyrmion deflection angle (see red and blue semicircles) is inverted for the cases $\xi<\alpha$ and $\xi>\alpha$.
Interestingly, the  velocity   of topologically non-trivial skyrmions depends on the  parameters  $\alpha$ and $\xi$ in rather different  ways. In the general case, this can be seen by inspecting \eqref{keymaster}, but we illustrate it  for the case of $\pi$-skyrmions  with  black dots and lines  in Fig.~\ref{big_circle_transform}\textbf{a} and \textbf{b}, respectively.
In particular, when $\alpha$  is constant, the trace of $(v_{\parallel}(\xi), v_{\perp}(\xi))$ is  a straight line for any topologically non-trivial skyrmion.  By contrast,   keeping  $\xi$ constant but varying $\alpha$ produces yet another circle for high-symmetry skyrmions. The trace of $(v_{\parallel}(\alpha), v_{\perp}(\alpha))$ is  a  section of the  circle  with equation
\begin{equation}
    \left(v_{\parallel}-\dfrac{1}{2}\right)^2 + \left(v_{\perp}-\dfrac{Q}{2\lambda\xi}\right)^2=\dfrac{1}{4}\left(1+\dfrac{Q^2}{\lambda^2 \xi^2}\right)
\end{equation}
where $\lambda=\lambda_{1}=\lambda_{2}$.
For the  $\pi$-skyrmion  ($Q=-1$) this is   the black circle shown in Fig.~\ref{big_circle_transform}\textbf{a}.
Note, the both cases $\alpha>\xi$ and $\alpha<\xi$ are realistic for different physical systems\cite{Malinowski}.

\begin{figure*}
\centering
\includegraphics[width=16.5cm]{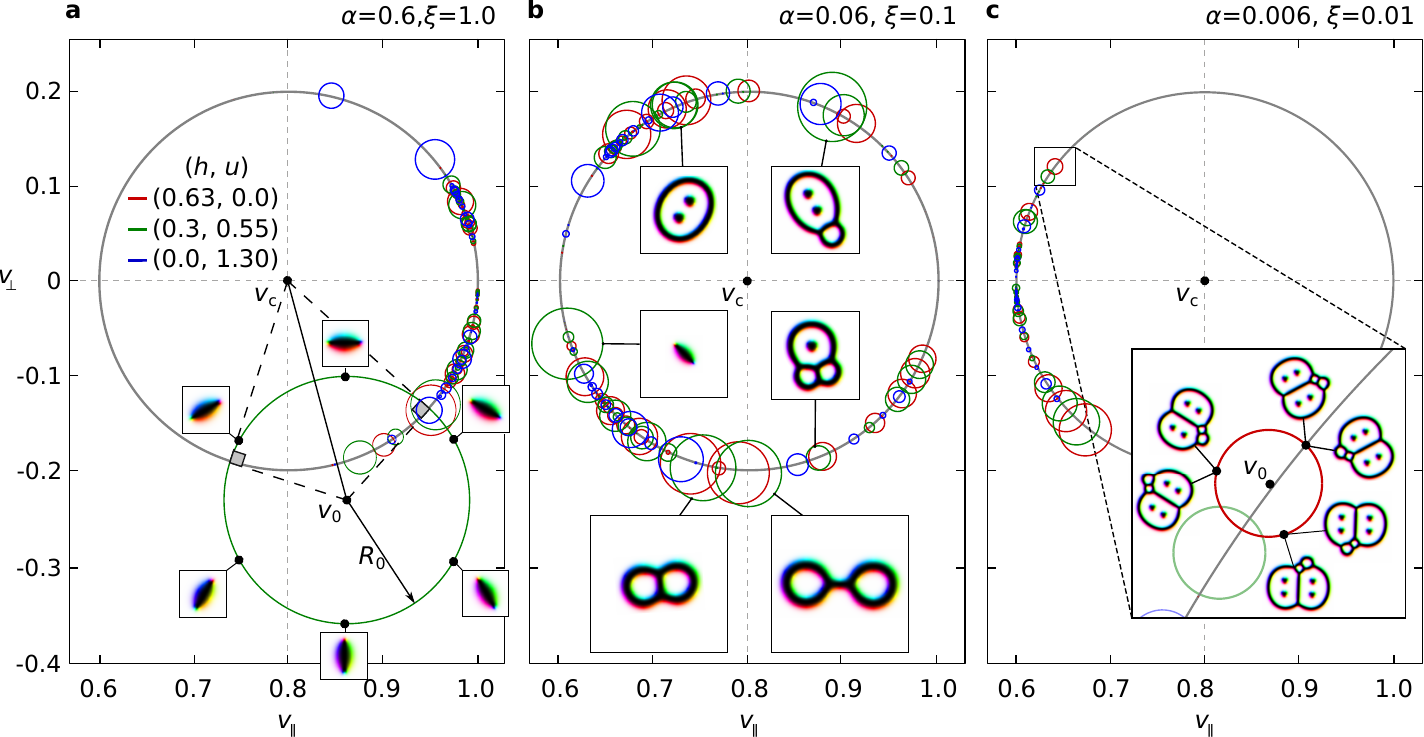}
\caption{\small \textbf{Velocity circles for low-symmetry skyrmions}. \textbf{a}-\textbf{c} show the  velocity distribution for various low-symmetry  skyrmions at different $\alpha$ and $\xi$. 
Some skyrmions are depicted in  the insets.
The grey velocity circle for high-symmetry skyrmions \eqref{big_circle} is provided for comparison.
The  velocity of low-symmetry skyrmions depends on the rotation angle of the skyrmion with respect to the current and forms an individual circle for each low-symmetry skyrmion as illustrated for the  antiskyrmion with $k_s=2$ in \textbf{a} and another low-symmetry skyrmion with $k_s=1$ in \textbf{c}.
}

\label{smal_circle}
\end{figure*}

\vspace{0.25cm}
\noindent
\textbf{Low-symmetry skyrmions}.
The class of low-symmetry skyrmions ($k_\mathrm{s}=1$ and 2) exhibits the most complicated but perhaps the most interesting geometrical properties among all three classes.
In particular, unlike high-symmetry skyrmions, the velocities of the low-symmetry skyrmions depend on the skyrmion's orientation relative to the direction of  the electric current.
The velocities corresponding to different rotation angles form a  circle in the velocity plane, uniquely determined  by the eigenvalues of the  skyrmion's dissipation tensor (Fig.~\ref{smal_circle}). These are the circles parameterised by the angle $\psi$ in \eqref{keymaster} at fixed $\vartheta$ and $\rho$.
Since the radii of these  circles are typically smaller than the radius $R_\mathrm{c}$ of the circle for high-symmetry skyrmions,  we refer to them as  small and large circles, respectively.
The insets in Fig.~\ref{smal_circle} \textbf{a} and \textbf{c} illustrate how the position on the small circle depends on the skyrmion rotation angle for a  skyrmion with $k_\mathrm{s}=2$ (antiskyrmion)  and for a  skyrmion with $k_\mathrm{s}=1$, respectively.
As seen from these figures, to make a single loop over the small circle, the low-symmetry skyrmion should be rotated by $\varphi=\pi$, as discussed   before equation \eqref{psicirclecentre}.
Under this rotation, the skyrmion of rotational symmetry $k_\mathrm{s}=2$ transforms into itself, and thereby each point on the circle corresponds to a unique configuration (Fig.~\ref{smal_circle} \textbf{a}).
In contrast to this, for skyrmions with $k_\mathrm{s}=1$, each point on the circle corresponds to two orientations of the spin texture, which differ on rotation by angle $\varphi=\pi$ (Fig.~\ref{smal_circle} \textbf{c}).

As follows from \eqref{psicirclecentre}, \eqref{psicircleradius}
the radius of a small circle $R_{0}$ and the position of its centre $(v_{\parallel}^{0},v_{\perp}^{0})$ are linked to the parameters of the big circle via the  equation
\begin{align}
    (v_{\parallel}^{0}-v_\mathrm{c})^2 + (v_{\perp}^{0})^2 =  R_\mathrm{c}^2 + R_{0}^2,
    \label{R-R0}
\end{align}
which implies  that any small circle and the big circle always intersect at right angles, as illustrated in Fig.~\ref{smal_circle}\textbf{a}. The equation \eqref{R-R0} also implies that the  position of the centre of any small circle is  outside the big circle and approaches it with decreasing $R_0$. In the limit $R_0\rightarrow 0$,
small circles degenerate into points on the big circle.
 
Based on the numerical experiments, we see that $R_{0}\ll R_\mathrm{c}$ for the majority of the skyrmions considered here. 
This in particular explains why the velocity distribution in Fig.~\ref{Fig1}\textbf{b} has a ring-like shape.

Plotting the velocities of a  large number of low-symmetry skyrmions for different $h$ and $u$ together we see that they indeed represent a set of circles, as shown in  Fig.~\ref{smal_circle}.
Three pair of parameters $\left(\alpha,\xi\right)$ with a fixed ratio $\xi/\alpha=5/3$ are provided to illustrate  the induced transformation of the  velocities which leaves  the large circle  unchanged.
Qualitatively the results are similar to those for high-symmetry skyrmions~Fig.~\ref{big_circle}\textbf{c}.
For decreasing damping, $\alpha \rightarrow 0$, the centres of the small circles \eqref{psicirclecentre} tend to one side of the large circle, $v_{\parallel}^{0}\rightarrow v_\mathrm{c}-R_\mathrm{c}$ (Fig.~\ref{big_circle}\textbf{a}), while for increasing damping
they approach the opposite side of the large circle, $v_{\parallel}^{0}\rightarrow v_\mathrm{c}+R_\mathrm{c}$ (Fig.~\ref{big_circle}\textbf{c}), with $v_{\perp}^{0}\rightarrow 0$ in both limiting cases.
Moreover, in both cases the radius \eqref{psicircleradius} of  the small circles goes to zero. This decrease is slow for the antiskyrmion because of its elongated shape.

\vspace{0.25cm}
\noindent
% \section*{Discussions}
\textbf{Discussions}

\noindent
\textbf{Speed limits}.
It is natural to ask which skyrmion moves the fastest  at fixed dynamical parameters $(I,\alpha,\xi)$. 
For $\alpha<\xi$, the topologically trivial skyrmions  are good candidates since they  move with the speed $v=1$ bounding the speed of high-symmetry skyrmions.
On the other hand, if the radius   $R_0$  \eqref{psicircleradius} of the  small circle for low-symmetry skyrmions becomes  sufficiently large then,  according to the general solution of Thiele's equation \eqref{keymaster},   the speed of such a skyrmion will exceed $v=1$ for a suitably chosen orientation. 
So there is no theoretical reason to rule  out speeds which exceed $v=1$. 
Although most of the low-symmetry skyrmions studied here typically have $v<1$, we found that the antiskyrmion depicted in Fig.~\ref{FigRot} \textbf{a} can, for certain orientations,  move faster than   topologically trivial skyrmions.
The maximum speed of the antiskyrmion,  corresponding to the point in the circle  which is furthest  from the origin Fig.~\ref{smal_circle} \textbf{a},  indeed exceeds $v=1$.
These observations are confirmed by numerical experiments based on the LLG equation.

To understand what makes the antiskyrmion special in this context one can refer to \eqref{psicircleradius}. According to that formula,  the radius of the small circle $R_0$ is directly proportional to the difference $\lambda_1-\lambda_2$ of the eigenvalues of the dissipation tensor.
Therefore,   elongated solitons with $\lambda_1 \gg\lambda_2$ are good candidates for high-speed solitons.
For the  antiskyrmion, the ratio $\lambda_{1}/\lambda_{2}$ is of order 5. 
We found other  skyrmion solutions with $\lambda_{1}/\lambda_{2}>5$, which according to Thiele's equation can move even faster than the antiskyrmion.
Numerical experiments, however, show that   skyrmions of such elongated shapes move as rigid objects only at very low currents. For realistic currents, the  skyrmions change shape so that the assumptions of Thiele's approximation, where the conservation of the skyrmion shape is essential, no longer hold. 

\vspace{0.25cm}
\noindent
\textbf{Materials with $D_{2d}$ symmetry}. 
The Lifshitz invariant in crystals with point group $D_{2d}$ has rotational symmetry~\cite{Bogdanov_89} like the invariants in Neel and Bloch-type systems.
In this case, however, the rotation in spin space should have the opposite  direction to  that given in \eqref{Rn}.
Since the dispersion tensor $\Gamma$ is invariant under rotations in spin space, the transformation rule \eqref{trS} of  $\Gamma$ is unchanged and   the   results of our theoretical analysis are fully applicable for solitons in these systems~\cite{Nayak_17} when accompanied by the exchange of particles and anti-particles ($Q \rightarrow -Q$).  

\vspace{0.25cm}
\noindent
\textbf{Other magnetic systems including centrosymmetric}. 
Chiral magnets possess a lot of similarities with other magnetic materials, for example  frustrated magnets~\cite{Leonov_15} and magnetic bubble materials~\cite{Malozemoff_79} -- films of centrosymmetric magnets with easy-axis perpendicular anisotropy. 
Although the mechanism for stabilization of magnetic solitons in these systems is quite different, the ground state (spirals or stripe domains) and the behavior of the system in an external field (the transition to a skyrmion lattice or bubble domain lattice) are very similar. 
The analysis of Thiele's equation presented here is independent of the  underlying 2D Hamiltonian. As a result, our classification of skyrmions into high-symmetry, low-symmetry and topologically trivial can be applied  to predict the  dynamics of any magnetic soliton responding to a current through the Zhang-Li torque.
 
Accordingly, in all such  systems,  topological solitons of high symmetry have velocities lying on a circle,  while  topologically trivial solitons  all have the same velocity $\mathbf{V}_0$. 
The theory developed here also predicts some properties of low-symmetry solitons. 
In particular, we expect  the rotation of low-symmetry solitons, if the Hamiltonian admits solitons degenerate in energy related by rotation, to generate circles in the velocity plane.
The size of these circles will be proportional to the difference in eigenvalues of the dissipation tensor, which is a measure of the soliton's elongation. 
Generally, stable configurations do not have strongly  elongated shapes, so that their velocities  lie close to the large circle \eqref{vcircle}. Therefore we expect the velocity distribution to have the ring-like shape found here for most materials. 
Finally we note that in the case of magnetic bubbles, three-dimensionality is crucial, and therefore, the theory proposed here may not cover cases where the magnetization is strongly inhomogeneous throughout the film thickness.

\vspace{0.25cm}
\noindent
\textbf{Geometry in velocity space}. We have seen that the apparently simple Thiele equation \eqref{Thiele} captures  surprising  geometrical features of skyrmion dynamics. They    are  revealed by the general solution \eqref{keymaster} and confirmed by numerical simulations. The geometrical features provide links with several themes in two-dimensional geometry even though the mathematics is superficially very different. The large circle  of high-symmetry skyrmion velocities provides the most basic illustration of this point. The mapping of a line into a  circle is a standard feature of    M\"obius transformations of the complex plane, and  writing the formula \eqref{keymaster} for $Q\neq 0$ and  $\vartheta=0$ in terms of complex numbers gives just  such a M\"obius transformation of the  scale parameter $\lambda= \lambda_1 =\lambda_2$. The small circles generated by varying $\psi$ in \eqref{keymaster} when $\vartheta >0$ intersect the large circle at right angles. Circles with this property are geodesics in the Poincar\'e disk model of the hyperbolic plane. Identifying the boundary of the Poincar\'e disk with our large circle therefore leads to an unexpected connection between Thiele's equation \eqref{Thiele} and hyperbolic geometry. As a final example of an unforeseen geometrical fact we show in the Method section that the velocities of low-symmetry skyrmions (with $\vartheta\neq 0$) trace out an ellipse  in velocity space when their overall scale is varied. Remarkably, this ellipse has the same eccentricity and orientation as the ellipse defined by  the dispersion tensor $\Gamma$. While the large circle and the small circles can easily be seen in simulations of actual skyrmions, and may be observable experimentally, the ellipses in velocity space are difficult to realise since they correspond to a special set of low symmetric skyrmions whose dissipation tensors have a fixed rotation angle and eigenvalue ratio.

\vspace{0.25cm}
\noindent
% \section{Methods}
\textbf{Methods}

\noindent
\textbf{Micromagnetic simulations} were performed with mumax code\cite{Mumax} on a rectangular domain, shape $L_\mathrm{x}\! \times \! L_\mathrm{y}$ with periodic boundary conditions (PBC). 
In general, the interaction between the skyrmion instances because of PBC may change the dynamics of the skyrmions.
This effect becomes especially pronounced when the domain of simulation is so small that it affects the shape and thus the symmetry of the skyrmion. 
To diminish this effect as far as possible  we use large size domains $L_\mathrm{x}, L_\mathrm{y} \sim 10 L_\mathrm{D}$.

To improve the accuracy in the  LLG simulations, instead of the second-order finite-difference scheme used by default in mumax, we implemented a fourth-order scheme in the spirit of the approach suggested by Donahue and McMichael~\cite{Donahue}.
For details, see Supplementary Note 1, where we discuss various aspects of the  accuracy in micromagnetic simulations and provide the mumax script with the fourth-order finite-difference scheme implemented.

The skyrmion position can be traced using the approach  suggested in Ref.~\onlinecite{Papanicolaou}, which is based on the formula for the centre of mass of a non-uniform rod \sout{where} but with  the topological density -- the integrand in Eq. \eqref{Qint} or magnon density~\cite{Kosevich} as in Ref.~\onlinecite{Komineas} -- playing  the role of distributed mass.
In long-time dynamics, when the skyrmion can cross the boundary of the simulated domain with PBC multiple times, this approach needs to be adapted.
In particular, when the skyrmion comes near the boundary of the simulation domain and part of it appears on the opposite side of the simulated domain, this formula suggests that the skyrmion slows down and starts to move in the opposite direction.   
We suggest an alternative approach to calculate the centre of the skyrmion, which follows from the solution of the problem for the centre of mass of a non-uniform ring.
The skyrmion position, $\left(x_\mathrm{s}, y_\mathrm{s}\right)$ can be defined as follows
\begin{align}
    x_\mathrm{s}\!=\!\dfrac{L_\mathrm{x}}{2\pi}\tan^{-1}\!\dfrac{\int\!\mathcal{N}_\mathrm{y}\sin\left(2\pi x/L_\mathrm{x}\right)\mathrm{d}x}{\int\!\mathcal{N}_\mathrm{y}\cos\left(2\pi x/L_\mathrm{x}\right)\mathrm{d}x}+l_\mathrm{x}L_\mathrm{x},\\ \nonumber
    y_\mathrm{s}\!=\!\dfrac{L_\mathrm{y}}{2\pi}\tan^{-1}\!\dfrac{\int\!\mathcal{N}_\mathrm{x}\sin\left(2\pi y/L_\mathrm{y}\right)\mathrm{d}y}{\int\!\mathcal{N}_\mathrm{x}\cos\left(2\pi y/L_\mathrm{y}\right)\mathrm{d}y}\pm l_\mathrm{y}L_\mathrm{y},
\end{align}
where $\mathcal{N}_\mathrm{x}\equiv\mathcal{N}_\mathrm{x}(y)=\int(1-n_\mathrm{z})\mathrm{d}x$ and $\mathcal{N}_\mathrm{y}\equiv\mathcal{N}_\mathrm{y}(x)=\int(1-n_\mathrm{z})\mathrm{d}y$ are the magnon density averaged along $x$ and $y$, respectively.
The integer numbers $l_\mathrm{x}$ and $l_\mathrm{y}$ stand for how many times the skyrmion has crossed the domain boundary in the $x$ and $y$ directions, respectively.
The sign in front of $l_\mathrm{y}$ depends on whether the skyrmion crosses the boundary in the positive or negative direction of corresponding axis, which in turn depends on the sign for the deflection angle, $\beta$. Since in our setup (Fig.~\ref{Fig1}\textbf{a}), the skyrmions move along the positive $x$-axis, the sign in front of $l_\mathrm{x}$ is always positive.
The approach to trace the position of the soliton presented  here is  similar to that for calculating the centre of mass for a set of point masses that are distributed in an unbounded 2D environment presented in Ref.~\onlinecite{Bai_08}.

The initial spin configurations for various types of skyrmions were either manually crafted in Excalibur code~\cite{Excalibur} according to the method described in Ref.~\onlinecite{Rybakov_19} or constructed through analytical functions as in Ref.~\onlinecite{Kuchkin_20ii}.

\vspace{0.25cm}
\noindent
\textbf{Thiele's equation and its general solution}.
We derive the general solution \eqref{keymaster} of Thiele's equation \eqref{Thiele}. 
With the abbreviation 
\[
\Omega =\begin{pmatrix} 0  & -1 \\ 1 & \phantom{-} 0 \end{pmatrix},
\]
for the $\pi/2$-rotation in the plane, we  write Thiele's equation as  
\begin{equation}
\left(\Gamma +\frac Q \alpha  \Omega \right)\mathbf{V}=\left(\Gamma +\frac Q \xi \Omega \right) \left(-\frac\xi \alpha\mathbf{I}\right).
\label{Thieleagain}
\end{equation}
Starting with the elementary observation that 
\[
\Gamma +\frac Q \xi\Omega = \frac{\xi+\alpha}{2\xi} \left(\Gamma +\frac Q\alpha \Omega \right) +  \frac{\xi-\alpha}{2\xi}  \left(\Gamma -\frac Q\alpha\Omega \right), 
\]
we deduce 
\begin{align*}
&\left(\Gamma +\frac Q\alpha\Omega  \right)^{-1}\left(\Gamma +\frac Q \xi \Omega \right)\\
&= \frac{\xi+\alpha}{2\xi}  + \frac{\xi-\alpha}{2\xi} 
\left(\Gamma +\frac Q\alpha \Omega \right)^{-1}\left(\Gamma -\frac Q \alpha\Omega \right).
\end{align*}
With $\vartheta = \ln\sqrt{\lambda_1/\lambda_2}$ as in the main text, we introduce the diagonal matrix 
\[
D(\vartheta) =   \begin{pmatrix} e^\vartheta & 0 \\ 0 & e^{-\vartheta }\end{pmatrix} ,
\]
and the rotation matrix 
\[
\qquad R(\varphi)= \begin{pmatrix}
\cos\varphi & -\sin \varphi \\ \sin\varphi & \phantom{-} \cos\varphi \end{pmatrix}
\]
we write as dissipation tensor as 
\[
\Gamma =  R(\psi) \begin{pmatrix} \lambda_1 & 0 \\ 0 & \lambda_2 \end{pmatrix} 
R(-\psi)=
 \sqrt{\lambda_1\lambda_2} R(\psi) D(\vartheta)R(-\psi),
 \]
to deduce, with  $\tan\dfrac \rho 2= \dfrac{Q}{\alpha \sqrt{\lambda_1\lambda_2}}$ as in the main text,  that 
\begin{align*}
&\left(\Gamma +\frac Q\alpha \Omega  \right)^{-1}\left(\Gamma -\frac Q \alpha \Omega \right)\\
& = R(\psi)\left(\Omega ^{-1} D(\vartheta)  +\tan\frac\rho 2 \right)^{-1}\left(\Omega^{-1}D(\vartheta) -\tan \frac \rho 2 \right)R(-\psi). 
\end{align*}
Noting that $ \Omega^{-1} D(\vartheta)\Omega^{-1} D(\vartheta)= -1 $
implies 
\[
\cos^2\frac \Phi 2\left(\Omega^{-1} D(\vartheta)  +\tan\frac\rho 2  \right)\left(\Omega^{-1} D(\vartheta)  -\tan\frac\rho 2  \right) =-1,
\]
and using trigonometric identities we  conclude 
\begin{align*}
&\left(\Gamma +\frac Q\alpha \Omega  \right)^{-1}\left(\Gamma -\frac Q \alpha \Omega \right)\\
& = \cos\rho - \sin\rho  \, \Omega R(\psi)D(\vartheta) R(-\psi) .
\end{align*}
Now multiplying out matrices and using  the  matrix 
\begin{equation}
P=\begin{pmatrix} 1 &\phantom{-} 0 \\ 0  & -1\end{pmatrix}
\label{Pmatrix}
\end{equation}
for the reflection on the $x$-axis, 
we arrive at the expression
\begin{align*}
& \left(\Gamma +\frac Q\alpha \Omega  \right)^{-1}\left(\Gamma -\frac Q \alpha \Omega \right)\\
&= \cos\rho - \sin\rho  \, \Omega 
\left(\cosh \vartheta \text{id}  + \sinh \vartheta R(2\psi)P\right).
\end{align*}
Using this to solve \eqref{Thieleagain} for $\mathbf{V}$, switching to $\mathbf{v}$ and expressing it   in terms of  the orthonormal basis $(\mathbf{e}_\parallel,\mathbf{e}_\perp)$ and  the   circle parameters \eqref{circleparameters}, one arrives at  the formula \eqref{keymaster}  in the main text. 

\vspace{0.25cm}
\noindent
\textbf{Circles and ellipses}.
We prove the geometrical results relating to  circles and ellipses in velocity space  discussed in the main text.
We write 
$P_\psi=R(2\psi)P= R(\psi)P R(-\psi)$ for the  the reflection on the line with  polar angle  $\psi$.
Denoting the polar coordinate of the direction $\mathbf{e}_\parallel$ by $\psi_\parallel$,  and with  $\psi'= \psi-\psi_\parallel$, we then have 
\[
P_\psi \mathbf{e}_\perp = \sin(2\psi')  \mathbf{e}_\parallel -\cos(2\psi')  \mathbf{e}_\perp,
\]
and so we can express the components of the velocity $\mathbf{v}$ in \eqref{keymaster}  also as
\begin{align}
v_\parallel-v_c&=R_c ( \cos\rho  +\sinh \vartheta \sin(2\psi')\sin \rho),\nonumber \\
 v_\perp& = -  R_c \sin\rho ( \cosh\vartheta + \sinh\vartheta \cos(2\psi')).
 \label{vpaperp}
\end{align}
We can use this formula to determine the  orientation of a low-symmetry skyrmion relative to the current  for which the response velocity $\mathbf{v}$ has the largest component $v_\parallel$. This is interesting since the angle $\psi$ can be changed by physically rotating a skyrmion. From \eqref{vpaperp}, we deduce  that, for $Q>0$,   $v_\parallel$ is maximal when $\psi'=\pi/4$ while for $Q<0$ it is maximal when   $\psi' =3\pi/4$. 

Equation \eqref{vpaperp} can also be used to show that the response velocities of skyrmions  with fixed  $\vartheta>0 $ and $\psi$ but different values of $\rho$ necessarily lie on an ellipse. 
One checks that the  velocity  satisfies the equation of an ellipse 
\[
(v_\parallel-v_c, v_\perp)M \begin{pmatrix} v_\parallel-v_c  \\ v_\perp\end{pmatrix} = R_c^2,
\]
 in terms of a symmetric and positive definite  matrix $M$ which is determined by $\vartheta$ and $\psi$  and which  is conveniently expressed in terms of its diagonal form as 
\begin{align*}
M = R(\psi) 
\begin{pmatrix} \lambda_+ & 0 \\ 0 &\lambda_- \end{pmatrix} R(-\psi),
\end{align*}
with eigenvalues proportional to those of $\Gamma$. In terms of the proportionality factor
$\kappa= \lambda_1\cos^2\psi' + \lambda_2 \sin^2 \psi'$, they are 
\begin{align*}
\lambda_+ = \lambda_1/\kappa,  \quad \lambda_- = \lambda_2/\kappa.
\end{align*}
The lengths of the  major and minor axes are therefore 
\begin{equation}
a =R_c/\sqrt{\lambda_-}, \quad
b =  R_c/\sqrt{\lambda_+},
\label{ellpara}
\end{equation}
giving the eccentricity 
\begin{equation} 
\varepsilon=  \sqrt{\frac{a^2-b^2}{a^2}} = \sqrt{\frac{\lambda_1-\lambda_2}{\lambda_1}}.
\label{eccentricity}
\end{equation}
The directions of the axes of the ellipse  are determined by the orthonormal basis which  also diagonalises the dissipation tensor, namely
\begin{equation}
\mathbf{e}_a=\begin{pmatrix} -\sin \psi  \\ \phantom{-} \cos \psi \end{pmatrix} ,\quad \mathbf{e}_b= \begin{pmatrix} \cos\psi  \\ \sin\psi \end{pmatrix}.
\label{ellaxes}
\end{equation}
Note that the major axis is in the direction of the eigenvector for the  smaller eigenvalue. The ellipses with these axes and ellipse parameters \eqref{ellpara} all go through the points $(\alpha/\xi) \mathbf{e}_\parallel$ and $\mathbf{e}_\parallel$, as can also be seen from \eqref{vpaperp}.

\vspace{0.25cm}
\noindent
\textbf{Acknowledgments}

\noindent
The authors acknowledge financial support from the European Research Council (ERC) under the European Union's Horizon 2020 research and innovation program (Grant No.\ 856538, project "3D MAGiC"), Deutsche Forschungsgemeinschaft (DFG) through SPP 2137 "Skyrmionics" Grant Nos.\ KI 2078/1-1 and SB 444/16. The work of K.\ Ch.\ was supported by 5-100 Russian Academic Excellence Project at Immanuel Kant Baltic Federal University. B.\ B.-S.\ acknowledges an EPSRC-funded PhD studentship.
F.N.R.\ acknowledges support from the Swedish Research Council Grants No.\ 642-2013-7837, 2016-06122, 2018-03659, and G\"{o}ran Gustafsson Foundation for Research in Natural Sciences.

\vspace{0.25cm}
\noindent
\textbf{Author contributions}

\noindent
V.M.K.\ and N.S.K.\ conceived the project, K. Ch.\ performed micromagnetic simulations with assistance of V.M.K.\ and N.S.K.\
V.M.K., B.B.-S.\ and B.J.S.\ performed the mathematical analysis of Thiele's equation.
All of the authors discussed the results and contributed to the writing of the manuscript.


\begin{thebibliography}{99}

\bibitem{Bogdanov_89}
Bogdanov, A.~N. \& Yablonskii, D. A. 
Thermodynamically stable ``vortices'' in magnetically ordered crystals. The mixed state of magnets. 
\href{http://www.jetp.ac.ru/cgi-bin/e/index/e/68/1/p101?a=list} {Sov. Phys. JETP \textbf{68}, 101 (1989)}.

\bibitem{Bogdanov_1994}
Bogdanov, A.~N. \& Hubert, A. 
The Properties of Isolated Magnetic Vortices,
\href{http://doi.org/10.1002/pssb.2221860223} {Phys. Status Solidi B \textbf{186}, 527 (1994)}.

\bibitem{Bogdanov_1994JMMM}
Bogdanov, A.~N. \& Hubert, A. 
Thermodynamically stable magnetic vortex states in magnetic crystals. 
\href{http://doi.org/10.1016/0304-8853(94)90046-9} {J. Magn. Magn. Mater. \textbf{138}, 255 (1994)}. 

\bibitem{Bogdanov_99}
Bogdanov, A.~N. \& Hubert, A. The stability of vortex-like structures in uniaxial ferromagnets. J. Mag. Mag. Mat. \textbf{195}, 182-192 (1999).

\bibitem{Rybakov_19}
Rybakov, F.~N. \&  Kiselev, N.~S.
Chiral magnetic skyrmions with arbitrary topological charge. \textit{Phys. Rev. B} \textbf{99}, 064437 (2019).

\bibitem{Foster_19}
Foster, D. et al.
% Foster, D., Kind, C., Ackerman, P.~J., Tai, J.-S. B., Dennis, M.~R. \& Smalyukh, I.~I.
Two-dimensional skyrmion bags in liquid crystals and ferromagnets.
{Nat. Phys. \textbf{15}, 655 (2019)}.

\bibitem{Kuchkin_20i}
Kuchkin, V.~M. \& Kiselev, N.~S. 
Turning a chiral skyrmion inside out.
\href{https://doi.org/10.1103/PhysRevB.101.064408}{Phys. Rev. B \textbf{101}, 064408 (2020).}

\bibitem{Kuchkin_20ii}
Kuchkin, V.~M. et al.
% Kuchkin, V.~M., B~Barton-Singer, F.~N. Rybakov, S.  Bl\"ugel, B.~J.  Schroers and N.~S. Kiselev
Magnetic skyrmions, chiral kinks and holomorphic functions.
\href{https://link.aps.org/doi/10.1103/PhysRevB.102.144422}{Phys. Rev. B \textbf{102}, 144422 (2020)}.

\bibitem{Kind_21}
Kind, C. \& Foster, D. 
Magnetic skyrmion binning,  \href{https://journals.aps.org/prb/pdf/10.1103/PhysRevB.103.L100413}{Phys. Rev. B. \textbf{103}, L100413 (2021)}.

\bibitem{Zeng_20}
Zeng, Z. et al. 
Dynamics of skyrmion bags driven by the spin–orbit torque.
\href{https://doi.org/10.1063/5.0022527}{Appl. Phys. Lett. \textbf{117}, 172404 (2020)}.

\bibitem{ZhangLi}
Zhang, S. \& Li, Z.
Roles of nonequilibrium conduction electrons on the magnetization dynamics of ferromagnets.
\href{https://doi.org/10.1103/PhysRevLett.93.127204} {Phys. Rev. Lett. \textbf{93}(12), 127204 (2004)}.


\bibitem{Dzyaloshinskii}
Dzyaloshinsky, I.
A thermodynamic theory of ``weak'' ferromagnetism of antiferromagnetics.
\href{http://doi.org/10.1016/0022-3697(58)90076-3} {J. Phys. Chem. Solids \textbf{4}, 241 (1958)}.

\bibitem{Moriya}
Moriya, T.
Anisotropic superexchange interaction and weak ferromagnetism. 
\href{http://doi.org/10.1103/PhysRev.120.91} {Phys. Rev. \textbf{120}, 91 (1960)}.


\bibitem{Romming_13}
Romming, N. et al.
Writing and Deleting Single Magnetic Skyrmions.
\href{http://doi.org/10.1126/science.1240573} {Science \textbf{341}, 636 (2013)}.

\bibitem{Kez_15}
K\'{e}zsm\'{a}rki, I. et al.
N\'{e}el-type skyrmion lattice with confined orientation in the polar magnetic semiconductor GaV4S8.
\href{http://doi.org/10.1038/nmat4402} {Nat. Mater. \textbf{14}, 1116 (2015)}.

\bibitem{Romming_15}
Romming, N., Kubetzka, A., Hanneken, C., von~Bergmann,  K. \& Wiesendanger, R. 
Field-Dependent Size and Shape of Single Magnetic Skyrmions. 
\href{http://doi.org/10.1103/PhysRevLett.114.177203} {Phys. Rev. Lett. \textbf{114}, 177203 (2015)}.

\bibitem{Nayak_17}
Nayak, A.~K. et al.
Magnetic antiskyrmions above room temperature in tetragonal Heusler materials.
\href{http://doi.org/10.1038/nature23466} {Nature \textbf{548}, 561 (2017)}.

\bibitem{ElisaGiovanni}
Davoli, E., Di Fratta, G., Praetorius D. \& Ruggeri, M. 
Micromagnetics of thin films in the presence of Dzyaloshinskii-Moriya interaction.
Preprint at \href{https://arxiv.org/abs/2010.15541} (2020).


\bibitem{Landau_Lifshitz}
Landau, L.~D. \& Lifshitz, E.~M. On the theory of the dispersion of magnetic permeability in ferromagnetic bodies. {Physik. Zeits. Sowjetunion \textbf{8}, 153 (1935)}.

\bibitem{Malinowski}
Malinowski, G., Boulle, O. \&  Kl\"{a}ui, M.
Current-induced domain wall motion in nanoscale ferromagnetic elements. \href{https://doi.org/10.1088/0022-3727/44/38/384005}{Journal of Physics D: Applied Physics \textbf{44}, 38 (2011)}.

% Published 8 September 2011 • 2011 IOP Publishing Ltd
% Journal of Physics D: Applied Physics, Volume 44, Number 38
% Citation G Malinowski et al 2011 J. Phys. D: Appl. Phys. 44 384005

\bibitem{Thiele_73}
Thiele, A.~A. Steady-State Motion of Magnetic Domains. {Phys. Rev. Lett. \textbf{30}, 230 (1973)}.

\bibitem{Komineas}
Komineas, S. \& Papanicolaou, N. Skyrmion dynamics in chiral ferromagnets under spin-transfer torque.
\href{http://dx.doi.org/10.1103/PhysRevB.92.174405} {Phys. Rev. B 92, 174405 (2015)}.

\bibitem{Q_note}
To avoid ambiguity, in definition of topological charge (\ref{Qint}) we follow the sign convention, see Ref.~\onlinecite{Melcher_14}, assuming that the skyrmion solutions always obey the condition $\mathbf{n}_0\!=\!(0,0,1)$ for $|\mathbf{r}|\!\rightarrow\!\infty$.

\bibitem{Malozemoff_79}{
Malozemoff, A.~P. \& Slonczewski, J.~C.
Magnetic Domain Walls in Bubble Materials (Academic Press, New York, 1979).}

\bibitem{Malozemoff_note}
Note, the definition for dissipation tensor in \eqref{dissipation} up to the factor $4\pi\alpha$ is equivalent to the definition of ``dissipation dyadic'' provided in Ref.~\onlinecite{Malozemoff_79}, see section VI.12 F Gyrovector and Dissipation Dyadic.

\bibitem{Barton-Singer_20} 
Barton-Singer, B., Ross, C. \& Schroers, B.~J. 
Magnetic skyrmions at critical coupling.
\href{https://doi.org/10.1007/s00220-019-03676-1} {Commun. Math. Phys. \textbf{375} 2259  (2020)}.

\bibitem{Excalibur}
Rybakov, F.~N. and Babaev, E. Excalibur software, 
\href{http://quantumandclassical.com/excalibur/} {\texttt{http://quantumandclassical.com/excalibur/}}.

\bibitem{SisodiaKomineas}
Sisodia, N., Muduli, P.~K., Papanicolaou, N. \& Komineas, S.  
Chiral droplets and current-driven motion in ferromagnets.
\href{https://doi.org/10.1103/PhysRevB.103.024431}{Phys. Rev. B \textbf{103}, 024431 (2021)}.

\bibitem{Leonov_15}
Leonov, A.~O. \& Mostovoy, M.
Multiply periodic states and isolated skyrmions in an anisotropic frustrated magnet.
\href{http://doi.org/10.1038/ncomms9275}{
Nat. Commun. \textbf{6} 8275 (2015)}.


\bibitem{Donahue}
Donahue, M.~J. \& McMichael, R.~D.  
Exchange energy representations in computational micromagnetics.
\href{https://doi.org/10.1016/S0921-4526(97)00310-4} {Physica B: Condensed Matter \textbf{223}, 4, 272-278 (1997)}.

% \bibitem{Wang_20}
% Wang, J. et al. Magnetic skyrmionium diode with a magnetic anisotropy voltage gating.
% \href{https://doi.org/10.1063/5.0025124}{
% Appl. Phys. Lett. \textbf{117}, 202401 (2020).}

% \bibitem{Ishida_20}
% Ishida, Y. \& Kondo, K. Theoretical comparison between skyrmion and
% skyrmionium motions for spintronics applications. 
% \href{https://iopscience.iop.org/article/10.7567/1347-4065/ab5b6b/pdf}{Jpn. J. Appl. Phys. \textbf{59}, SGGI04 (2020).} 

\bibitem{Mumax}
Vansteenkiste, A. et al. 
The design and verification of MuMax3.
\href{https://doi.org/10.1063/1.4899186} {AIP Advances \textbf{4}, 10 (2014)}.


\bibitem{Papanicolaou}
Papanicolaou, N. \& Tomaras, T.~N.
Dynamics of magnetic vortices.
\href{https://doi.org/10.1016/0550-3213(91)90410-Y}{Nucl. Phys. B \textbf{360}, 425 (1991)}.

\bibitem{Kosevich}
Kosevich, A.M., Ivanov, B.A. \& Kovalev A.~S. Magnetic solitons.
\href{https://doi.org/10.1016/0370-1573(90)90130-T}{Physics Reports \textbf{194}, 117 (1990)}.

\bibitem{Bai_08}
Bai, L. \& Breen, D.
Calculating Center of Mass in an Unbounded 2D Environment.
\href{http://dx.doi.org/10.1080/2151237X.2008.10129266}{ J. Graph. Tools, \textbf{13}, 53 (2008)}.

% \bibitem{Hoffmann}
% Hoffmann, M. et al.
% Antiskyrmions stabilized at interfaces by anisotropic Dzyaloshinskii-Moriya interactions, 
% \href{http://doi.org/10.1038/s41467-017-00313-0} {Nat. Commun. \textbf{8}, 308 (2017)}. 




\bibitem{Melcher_14}
Melcher, C.
Chiral skyrmions in the plane.
\href{http://doi.org/10.1098/rspa.2014.0394} {Proc. R. Soc. A \textbf{470}, 20140394 (2014)}. 









\end{thebibliography}
\end{document}